\documentstyle[11pt,COIMBRA_pasp,epsf]{article}

\begin{document}

\title{Temperature predictions for dust embedded AGB stars: 
how far can we go with synthetic near-infrared photometry ? } 

\author{E. Lastennet, S. Lorenz-Martins, F. Cuisinier}
\affil{Depto. de Astronomia, UFRJ, Rio de Janeiro, Brazil}

\author{T. Lejeune}
\affil{Observat\'orio Astron\'omico, Coimbra, Portugal}

\begin{abstract}

We discuss the possibilities of the BaSeL models in its lowest 
temperature boundary (T$_{\rm eff}$$\sim$2500 K for cool giants) to 
provide the T$_{\rm eff}$ of AGB stars. 
We present the first step of our work, by comparing our predictions 
for the AGB star R Fornacis with the results of Lorenz-Martins \& 
Lef\`evre (1994) based on the dust spectral energy distribution. 

\end{abstract}

% Keywords should be included, but they are not printed in the hardcopy.


\section{Introduction}

Main sequence stars with mass in the range  0.9-9 M$_{\odot}$
evolve through a double shell burning phase, refered to as
the asymptotic giant branch (AGB) phase of evolution.
This phase is characterized by carbon dredge up of the core
to the surface after each thermal pulse - Helium shell flash -
(Iben \& Renzini 1983). 

The temperatures of these objects are very badly known. Although they are
highly variable, their determination from static models such as assumed in 
the BaSeL library can be justified as a first approximation.  
In order to explore the capabilities of the BaSeL library (Lejeune,
Cuisinier \& Buser 1997, 1998 and references therein, see also Lastennet,
Lejeune \&  Cuisinier, these proceedings) to predict correct temperatures for
such cool AGB  stars, we compare our results from synthetic infrared
photometry of the stellar  photosphere with the detailed study of
Lorenz-Martins \& Lef\`evre (1994)  of the AGB carbon star R Fornacis. Their
work is based on a modelling of the  spectral energy distribution of the
dust envelope, where they put tight  constraints on the temperature of the
heating source. 

\section{R Fornacis as a test case}

Table 1 gives the JHKLM photometry of R For (HIP 11582) that we used (Le
Bertre, 1992). The photometric errors in the individual JHKLM magnitudes are
not provided so we assume an error of 0.2 on each magnitude, according to the 
maximum uncertainty estimated from Fig. 1 of Le Bertre (1988).  

%\scriptsize 
\begin{table*}[hbt]
\caption[]{
Infrared JHKLM photometry (Le Bertre, 1992, Julian date 7643) and effective
temperature of the central star of R Fornacis.
}
\label{tab:data}
\begin{flushleft}
\begin{center}
%\scriptsize    
\begin{tabular}{ccccccc}
\tableline\noalign{\smallskip} 
  J  &  H  &  K  &  L  &  M  & T$_{\rm eff}$$^{(1)}$ &  T$_{\rm eff}$$^{(2)}$\\ 
      &     &     &     &     &   (K)          & (K) \\ 
\noalign{\smallskip}
\tableline\noalign{\smallskip} 
 5.76 & 3.97 & 2.32 &  0.21 & $-$0.28 &  2650 & 2440-2520 \\
\noalign{\smallskip}
\tableline
\tableline
\noalign{\smallskip}
\noalign{\smallskip}
\end{tabular}
\end{center}
%\\   
%\small
%\scriptsize 
$^{(1)}$ Lorenz-Martins \& Lef\`evre (1994); \\
$^{(2)}$ BaSeL JHKM synthetic
photometry (this work, see text for details). 
%\label{tab:data}
\end{flushleft} 
\end{table*}     

\normalsize

\section{Method} 

Although the dust may have a significant contribution in the IR {\it bands}  
of this star, especially L and M, it should only have a secondary 
influence on the photospheric {\it colours}.
We intend of course to correct for the predicted differences by a dust 
model (Lorenz-Martins \& Lef\`evre, 1993) due to the envelope. However
in a first step we merely  compare the observed colours of R Fornacis  
with the photospheric predictions of the BaSeL library 
(BaSel-2.2 version, with spectral corrections) by 
minimizing their $\chi^2$ differences. \\
This $\chi^2$-minimization method
is similar to the one applied in Lastennet et al. (2001): we derived the 
T$_{\rm eff}$ and log g  values matching simultaneously the 
observed JHKLM photometry listed in Tab. 1, assuming a solar metallicity
([Fe/H]$=$0).   

\section{Preliminary results}

We have tested various colour combinations of the J (1.25 $\micron$), 
H (1.65 $\micron$), K (2.2 $\micron$), L (3.4 $\micron$), 
and M (5.0 $\micron$) magnitudes: (J$-$H), (H$-$K),
(K$-$L), (J$-$K) and (K$-$M). They all give T$_{\rm eff}$ estimates in
agreement with the work of Lorenz-Martins \& Lef\`evre (1994). \\
Since better constraints should be obtained by matching more than 1 colour, 
we chose the (J$-$H) and (K$-$M) colours which give the best $\chi^2$-scores. 
The solutions we get to match simultaneously the observed (J$-$H) and (K$-$M)
are presented in Fig. 1. 
Our best BaSeL-infrared solution is T$_{\rm eff}$$=$2440K\footnote{Note: for
giants, BaSeL solutions cooler than 2500K are extrapolated.}, but all 
the solutions inside the 1-$\sigma$ contour are good fits to the observed 
photometric data.   
The effective temperature of the  central star of R For found by 
Lorenz-Martins \& Lef\`evre is T$_{\rm eff}$$=$2650 K 
(shown as a vertical line on Fig. 1). This is larger by $\sim$100K than the 
1-$\sigma$ BaSeL contour but still inside the 2-$\sigma$ contour.  
Additionally the BaSeL models show that this star has a surface gravity log g
$\sim$$-$0.5$\pm$0.4, which is  what one expects for carbon stars. 

\section{Conclusion}

We reported a preliminary study to determine the T$_{\rm eff}$ and surface 
gravity of the central star of R Fornacis by exploring the best $\chi^2$-fits
to the infrared photometric data. 
These results are in a surprising good agreement - given the 
approximation we made (no envelope absorption/emission correction) - 
with the detailed study of Lorenz-Martins \& Lef\`evre (1994). 
Therefore, while detailed spectra studies are obviously highly preferred 
(see e.g. Loidl, Lan\c{c}on \& J{\o}rgensen, 2001), our method 
may provide a good starting point. 
If our R Fornacis result is confirmed with other AGB stars, 
this would mean that the BaSeL JHKLM synthetic photometry 
is suited to derive (Teff-log g) estimates for cool AGB stars. 

%%%%%%%%%%%%%%%%%%%%
\begin{figure}
\centerline{    
\plotfiddle{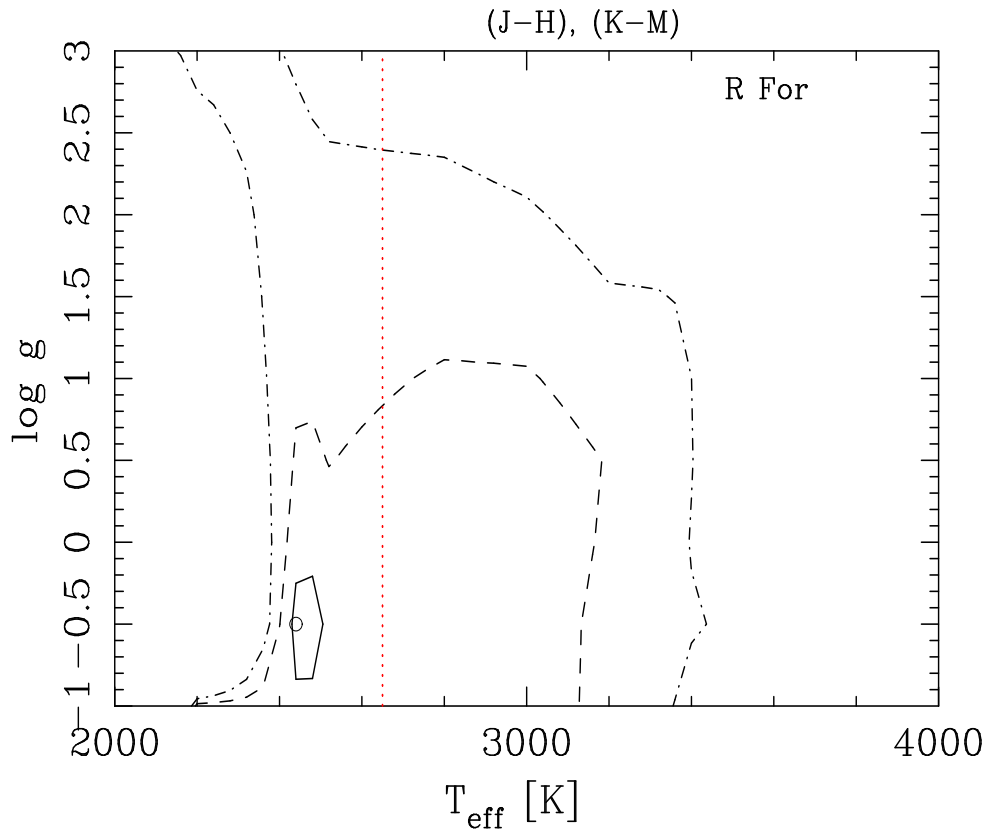}{11.5cm}{-90}{50}{55}{-400}{330}
%\plotfiddle{rfor.ps}{11.5cm}{-90}{120}{120}{-370}{360}
}
 \caption{
 \small 
Determination of the temperature and surface gravity 
of R Fornacis from the infrared synthetic photometry of the 
BaSeL models. The best solutions are inside the 1-$\sigma$ 
contours defined by a small region (solid line). 
The 2- (dashed line) and 3-$\sigma$ (dot-dashed 
lines) are also shown. The determination of Lorenz-Martins \& Lef\`evre 
(1994) is displaied as a vertical dotted line at T$_{\rm eff}$$=$2650 K for
comparison. 
\normalsize
 }
%\label{f:contours}
\end{figure}
%%%%%%%%%%%%%%%%%%%%

 \acknowledgments
EL acknowledges support from CNPq and FAPERJ, and   
thanks the conference organizers for a stimulating 
meeting and for financial support.

\end{document}